\documentclass[twocolumn,amssymb,prl]{revtex4-1}
\usepackage{graphicx,pgfplots}
\usepackage{dcolumn}
\usepackage{bm}
\usepackage{hyperref}
\usepackage{enumerate}
\usepackage{comment}

\begin{document}
\preprint{APS/123-QED}

\title{Disorder Driven Metal-Insulator Transition in BaPb$_{1-x}$Bi$_x$O$_3$ and Inference of Disorder-Free Critical Temperature}
\author{Katherine Luna, Paula Giraldo-Gallo, Theodore Geballe, Ian Fisher, Malcolm Beasley}
\affiliation{Department of Physics, Stanford University, Stanford CA 94305-4045, USA} 
\date{\today}

\begin{abstract}
We performed point-contact spectroscopy tunneling measurements on single crystal BaPb$_{1-x}$Bi$_x$O$_3$ for $x=0$, $0.19$, $0.25$, and $0.28$ at temperatures ranging from $T=2-40$ K and find a suppression in the density of states at low bias-voltages, which has not previously been reported.  The classic square root dependence observed in the density of states fits within the theoretical framework of disordered metals, and we find that the correlation gap disappears around a critical concentration $x_c=0.30$.  We also report a linear dependence of the zero-temperature conductivity, $\sigma_0$, with concentration, where $\sigma_0=0$ at $x_c=0.30$.  We conclude that a disorder driven metal-insulator transition occurs in this material before the onset of the charge disproportionated charge density wave insulator.  We explore the possibility of a scaling theory being applicable for this material.  In addition, we estimate the disorder-free critical temperature using theory developed by Belitz \cite{Belitz89} and Fukuyama et. al.\cite{Fukuyama84} and compare these results to Ba$_{1-x}$K$_x$BiO$_3$.
\end{abstract}

\pacs{74.45.+c}
\keywords{superconductivity, tunneling, disorder, proximity effect, point-contact spectroscopy}
\maketitle

The bismuthate superconductors (doped BaBiO$_3$) were the first class of  oxide superconductors to be discovered \cite{Sleight75, Cava88}. They exhibit moderately high superconducting transition temperatures (up to $\sim12$ K in BaPb$_{1-x}$Bi$_x$O$_3$ (BPBO) and $\sim 30$ K in BaK$_x$Bi$_{1-x}$O$_3$ (BKBO)), and they are another example of a high $T_c$ superconducting phase adjacent to a competing ordered phase, only in this case the ordered phase is in the charge sector \cite{Uchida87,Tarapheder96}. They were highly studied in the era before the discovery of the cuprate superconductors. 

 Still, despite this considerable effort, neither the electronic structure of these materials nor the ingredients of their superconductivity could be satisfactorily treated theoretically \cite{Meregalli98,Mattheiss83, Mattheiss85}.  Simple valance arguments suggest that the parent compound BaBiO$_3$ should be a half-filled band metal with Bi in a ${4+}$ valence state, whereas in fact it is an insulator due to charge disproportionation (e.g. Bi$^{4+}\rightarrow$ Bi$^{3+}+$Bi$^{5+}$) lending to a so-called charge disproportionated charge density wave (CD-CDW), which is a distinct form of CDW not associated with Fermi surface nesting.  One can also think of the CD-CDW as arising from a negative-U on the Bi sites.   Traditional density functional  electronic structure calculations were not able to account for this CD-CDW state, and the most up to date  calculations of the electron-phonon interaction parameter $\lambda$ yield values that are too small to account for the observed high transition temperatures \cite{Meregalli98}.

Recently the theoretical situation has greatly improved.  Franchini et. al.  first showed that the insulating state (as well as the structure and lattice constants) of BaBiO$_3$ could be understood within density functional theory if a new functional (the so-called HSE functional) was used \cite{Franchini09,Franchini10}.  This functional is computationally more complex but incorporates better the Coulomb correlations present in the bismuthates.  Using this approach, Yin et. al. showed that  $\lambda$ in the bismuthates was  ``dynamically" enhanced and  that these larger values could  account for the observed transition temperatures.  In their work, to calculate $T_c$, these authors used the strong-coupled McMillan formulation of the Eliashburg theory and their calculated values of $\lambda$ and the renormalized Coulomb interaction parameter $\mu^*$ \cite{Yin11,Yin13}.  

In this Letter, we show that the effects of disorder (localization) are another essential factor in understanding these materials that has not been appreciated  previously.  Specifically, we show that in BPBO there is a disorder-induced metal-insulator transition at a composition $x_c < x_{\textrm{CDW}}$ where $x_{\textrm{CDW}}$ is the critical concentration at which the CD-CDW state forms.  For $x<x_c$ we also observe a reduction in the tunneling density of states at the Fermi level that is expected due to electron-electron interactions in the presence of disorder.  When such disorder effects are present, one also expects
 a reduction of $T_c$  due to a disorder-enhanced $\mu^*$, as first noted by Fukuyama et. al. \cite{Fukuyama84}.   
 
 Building on this fact,  and using the most complete theory of the effects of disorder on  $T_c$, we  show that   it is possible to  back out  an estimate of the disorder-free transition temperature from our data.  The result is that in the case of BPBO the maximum inferred disorder-free $T_c$ is around a factor of 2 higher than the experimental value at optimal doping.  In the larger picture, we believe our results are relevant for determining the intrinsic $T_c$ in any material for which the resistivity (or for a film, the sheet resistance) is large enough that it is in the regime of localization effects and for which in the pairing interaction the Coulomb repulsion between the pairs is avoided by retardation.

\begin{figure}[t!]
  \centering
  \includegraphics[width=0.4\textwidth]{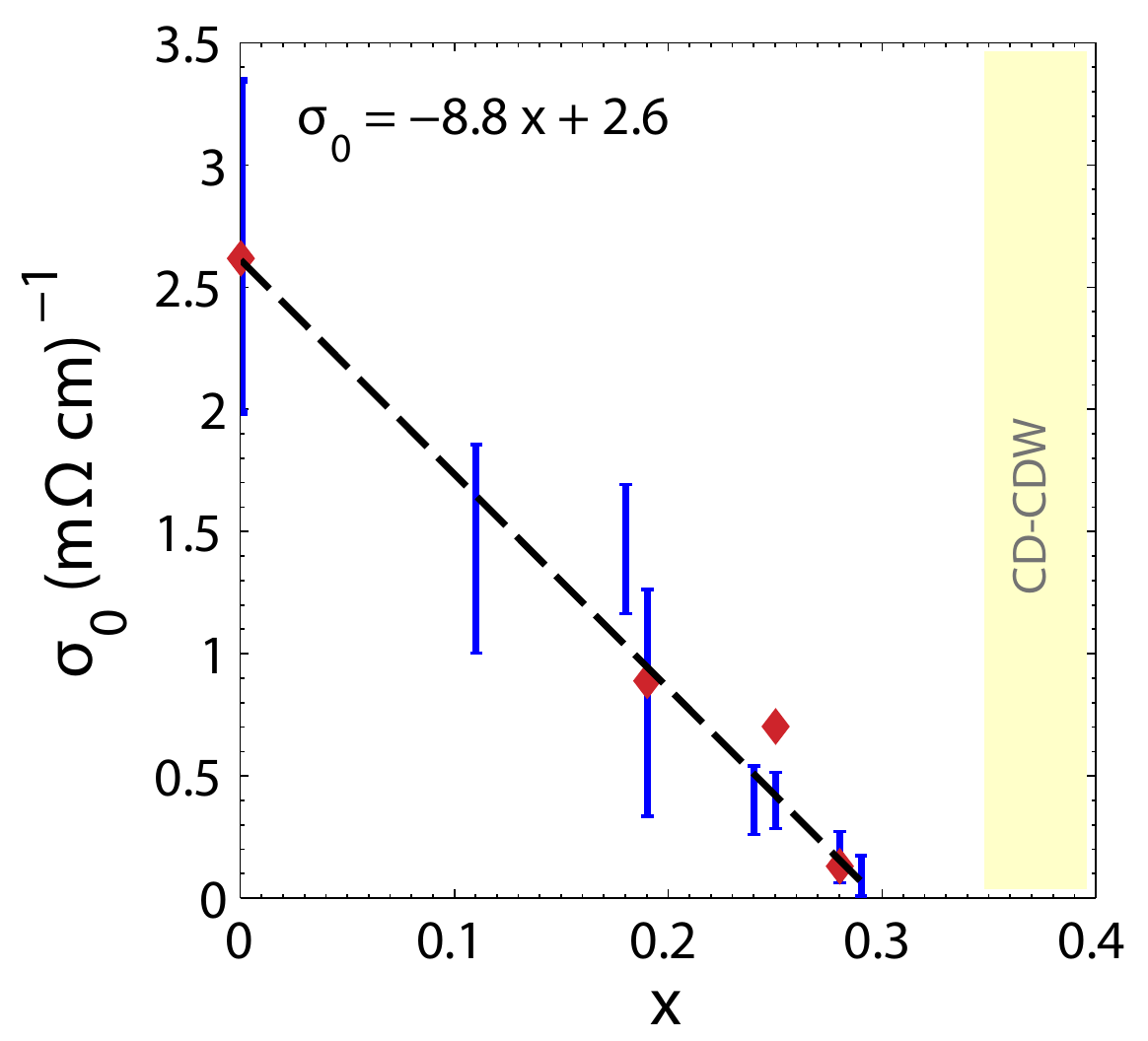}
  \caption{Zero temperature conductance vs. doping obtained from a four-point resistivity vs. temperature measurement. The zero bias conductivity is estimated from Ref. \cite{GiraldoGallo12} with variance due to geometrical factors from four to five crystals measured per concentration.  Red diamonds correspond to the zero temperature conductivity for the samples used in the tunneling experiments. Dashed line is a linear fit to these data points.}
\label{fig:cond0vsDoping}
\end{figure}

The existence of a metal-insulator transition is  demonstrated in Fig. \ref{fig:cond0vsDoping}  where the zero-temperature conductivity is plotted as a function of composition.  The conductivity decreases linearly to the critical value  $x_c$= 0.30  The blue lines are obtained from the four-point resistivity measurement shown in  Fig. 2 of Ref. \cite{GiraldoGallo12}, where a linear extrapolation is made using points prior to the onset of $T_c$.  The variation is due to geometrical factors from four to five resistivity measurements per doping concentration.  The red diamonds correspond to the zero temperature conductivity for the exact samples used in the tunneling measurements discussed below.  Note that in the literature the best estimates of the  concentration for the onset of the CD-CDW state is $x_{\textrm{CDW}}$= 0.35 \cite{Uchida87}. (See yellow region in the figure.)

\begin{figure}[t!]
  \centering
   \includegraphics[width=0.4\textwidth]{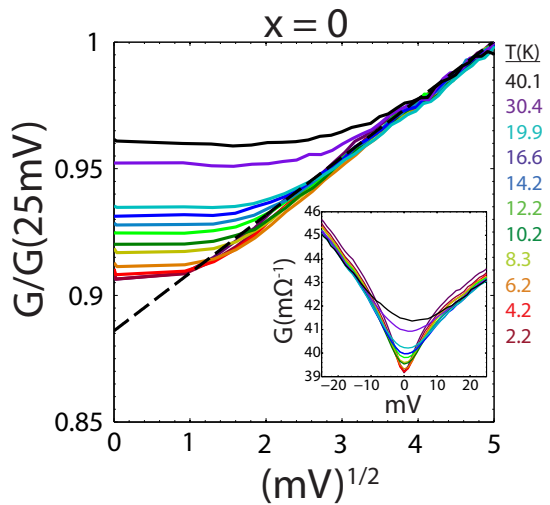}
  \caption{Differential conductance normalized at 25 mV as a function of the square root of positive bias-voltage for $x=0$ at various temperatures.  Inset shows the raw data of differential conductance as a function of voltage at low bias-voltages for various temperatures.}
\label{fig:Condsqrt_and_highBias}
\end{figure}

An example of our tunneling data are shown in Fig. \ref{fig:Condsqrt_and_highBias}.  The data were obtained using  point-contact spectroscopy (PCS) measurements on single-crystals of BaPb$_{1-x}$Bi$_{x}$O$_3$ (BPBO) with doping concentrations $x=0$, $0.19$, $0.25$, and $0.28$, and grown in a method described in Ref. \cite{GiraldoGallo12}.  Measurements were performed from temperatures ranging from $2-40$ K.  The junctions were prepared by cleaving the sample in air and then bringing the sample in contact with the tip at room temperature.  The apparatus was then inserted into a flow cryostat for measurements.  In the superconducting range of compositions,  superconducting densities of states were observed that are consistent with those reported in the past by many researchers \cite{Ekino89,Dynes91PRB, Szabo97, Suzuki92}.

However, in Fig.  \ref{fig:Condsqrt_and_highBias}, we focus on the behavior of the density of states above $T_c$, which is a region where little attention has been given.  A cusp is observed in the differential conductance measurement, as shown for example in the insert of Fig. \ref{fig:Condsqrt_and_highBias}, which shows the tunneling density of states at low bias-voltages for the sample with $x=0$ (i.e., BPO, which is not a superconductor).  Similar cusps are seen for all concentrations $(x \le 0.28)$ including those that are superconducting.   To our knowledge, this cusp has not been noted previously, where historically attention has been focused on the unexplained asymmetric v-shaped tunneling density of states at higher voltages \cite{Sharifi91}.  On the other hand, the cusp we report is similar to that seen in  amorphous Nb-Si alloys \cite{Hertel83}, which is one of the classic cases of  a disorder driven (localization) metal-insulator transition.  In plotting the data in Fig. \ref{fig:Condsqrt_and_highBias} the data have been normalized to the differential conductance at 25 mV, which we take as a measure of the background density of states free of disorder effects.  There is some arbitrariness in this choice due to the unexplained linear background at high bias voltages universally seen in bismuth tunneling data.  On the other hand, examination of the insert in the figure indicates that the zero-bias anomaly of interest to us merges into the linear background in the vicinity of 25 mV.

The theory of the reduction of the density of states $N(E)$  due to  disorder-enhanced Coulomb interactions  is well established.  In three dimensions, It predicts that $N(E)=N(0)[1+(E/\Delta)^{1/2}]$, where $N(E)$ is the density of states at zero temperature and $\Delta$ is the correlation gap \cite{Altshuler79}.  As shown in the main part of Fig. \ref{fig:Condsqrt_and_highBias}, our data follow this energy dependence very well.  Here we have plotted the normalized tunneling density of states versus the square root of the bias voltage for various temperatures.  From the fit to the data (dashed line in the figure), we can determine both the correlation gap $\Delta$ (inverse slope) and the zero-temperature reduction in the density of states at zero-bias $N(0)$ (zero voltage intercept).

\begin{figure}[t!]
  \centering
   \includegraphics[width=0.45\textwidth]{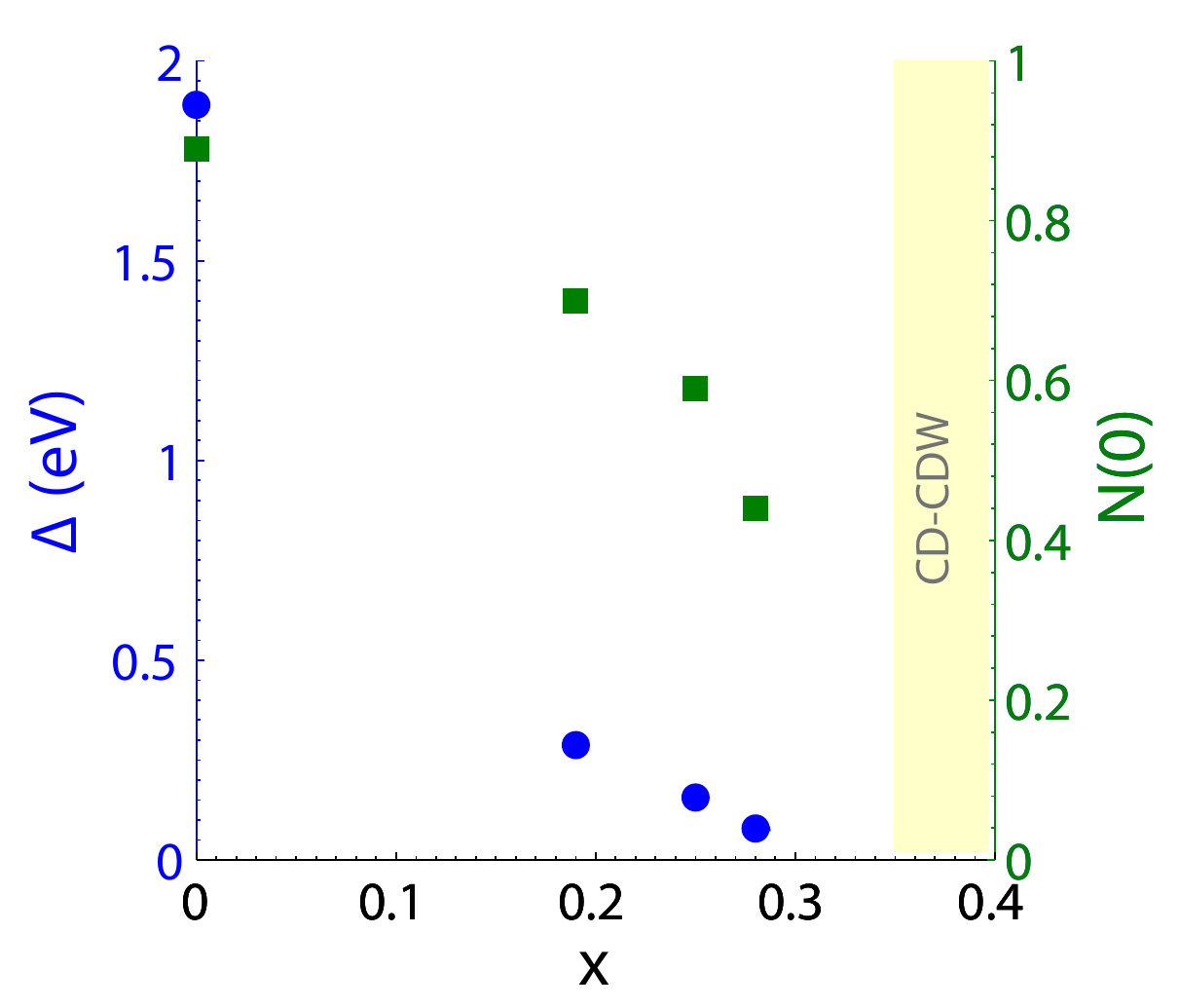}
  \caption{Correlation gap, $\Delta$ (blue circles), and density of states, N(0) (green squares), at zero temperature as a function of doping, $x$.  These quantities are extracted from differential conductance data, where for example in Fig. \ref{fig:Condsqrt_and_highBias}, $\Delta$ and N(0) correspond to the inverse of the slope and intercept respectively of the dashed line.  Results derived from the differential conductance at positive bias-voltage is shown. }
\label{fig:CorrGapN0_vs_x_negPosbias}
\end{figure}

 A similar procedure can be performed for the other concentrations, and the results are shown in Fig. \ref{fig:CorrGapN0_vs_x_negPosbias}.  As the conductance is asymmetric, results differ between positive and negative bias-voltages.  The differences are not large, however, and for clarity of presentation we show only the data for positive bias.   The blue circles represent the correlation gap, $\Delta$, and the green squares represent the density of states at zero temperature and zero-bias voltage, $N(0)$.    As is evident in the figure, the correlation gap nicely extrapolates to zero at $x_c$=0.30, as determined from the transport data.

Having established the existence of a metal-insulator transition due  to disorder, it is of interest to compare our results with McMillan's scaling theory \cite{McMillan81} of such transitions that was developed to account for the disorder-driven metal-insulator transition seen in Nb$_x$Si$_x$ \cite{Hertel83}.    The scaling theory involves two critical exponents $\nu\sim1$ and $1<\eta<3$.  In terms of these exponents, the theory predicts for $E<\Delta$ that  $\sigma_0\sim(x-x_c)^{\nu}$ , $N(E)=N(0)[1+(E/\Delta)^{1/2}]$, $\Delta=(x-x_c)^{\nu\eta}$ and $N(0)\sim(x-x_c)^{\nu(3-\eta)}$.  Our transport and tunneling density of states data are nicely consistent with the first two predictions of the theory and yield a value $\eta=1$.  The fits for $\Delta$ and $N(0)$ as functions of $x$ are less satisfactory. The first yields $\eta=1.7$ and the second, $\eta=2.7$.

Granted more data points would yield  more accurate results. Also, as noted above, some uncertainty can be associated with the normalization procedure use in the tunneling data.  In addition, we should note that the scaling theory is only valid around the critical region whereas we are including points at $x=0$, which is relatively far from the critical concentration $x_c$.  And, as pointed out by Lee et. al. \cite{Lee82,Lee85}, while McMillan's scaling theory is a good starting point for data analysis, there are some subtleties in his assumptions about the screening constant and conductivity being proportional to the single particle density of states, $N(0)$, rather than the change of density with chemical potential, $\textrm{d}n/\textrm{d}\mu$.  Last but not least, the theory does not consider what would happen when the metal-insulator transition is very near a CD-CDW transition.   In short, we are entering unexplored territory.


Let us now turn to the issue of the reduction of $T_c$ due to disorder. From the work of Belitz \cite{Belitz89}, which is the latest word on this subject for three dimensional systems, we have a McMillan-like equation for $T_c$ valid for strong coupling and relatively strong disorder.  

\begin{equation}
T_c = \frac{\Theta_{D}}{1.45}\textrm{exp}\left[\frac{-1.04(1+\tilde{\lambda}+ Y')}{\tilde{\lambda} - \tilde{\mu}^*[1+0.62\tilde{\lambda}/(1+Y')]}\right].
\label{TcBelitz}
\end{equation}
The actual theory uses the prefactor $\omega_{\textrm{log}}/1.2$ in front of the exponential, but as not enough information is known to determine $\omega_{\textrm{log}}$, we fall back to an older form of the McMillan formulation. 

Conveniently for our purposes,  the disorder is parameterized  by the fractional reduction of the density of states at the Fermi energy.
 
\begin{equation}
Y' = N_F/N(\bar{\omega})-1
\end{equation}
where $N(\bar{\omega})$ is the density of states evaluated at a characteristic phonon frequency, and $N_F$ is the clean normal-metal DOS at the Fermi level.  For simplicity we have taken $\bar{\omega}=0$.
 
 $Y'$ enters the equation for the reduction of $T_c$ both explicitly as shown in the equation  and implicitly through the disorder-dependent electron-phonon coupling $\tilde{\lambda}(Y')>\lambda$ and the disorder-dependent Coulomb pseudopotential $\tilde{\mu}^*(Y')>\mu^*$.

 In the theory of Belitz, both $\tilde{\mu}^*$ and $\tilde{\lambda}$ also depend on the ratio between the Thomas Fermi screening wave number and the Fermi wavenumber, $x=2k_F/k_{TF}$.  We estimate these wave numbers using simple band relations $k_F=(3\pi^2n)^{1/3}$ and $k_{TF} = (6\pi n e^2/\epsilon_{\infty} E_F)^{1/2}$, where $\epsilon_{\infty}=1$.  Experimental results of the Debye temperature \cite{Itoh84}, the Fermi energy \cite{Tani80}, and the carrier density \cite{Thanh80} were used where there is a nice summary of these parameters for various concentrations shown in Table 1 of Ref. \cite{Kitazawa85}.  The renormalized Coulomb interaction $\mu^*$ with no disorder is also estimated using the Morel-Anderson equation, $\mu^*=\mu/[1+\mu\textrm{ln}(E_F/k_B\Theta_D)]$ with $\mu=(1/2x^2)\textrm{ln}(1+x^2)$ \cite{Morel62}.
This procedure produces the value $\mu^*$ = 0.14.

\begin{figure}[t!]
  \centering
   \includegraphics[width=0.45\textwidth]{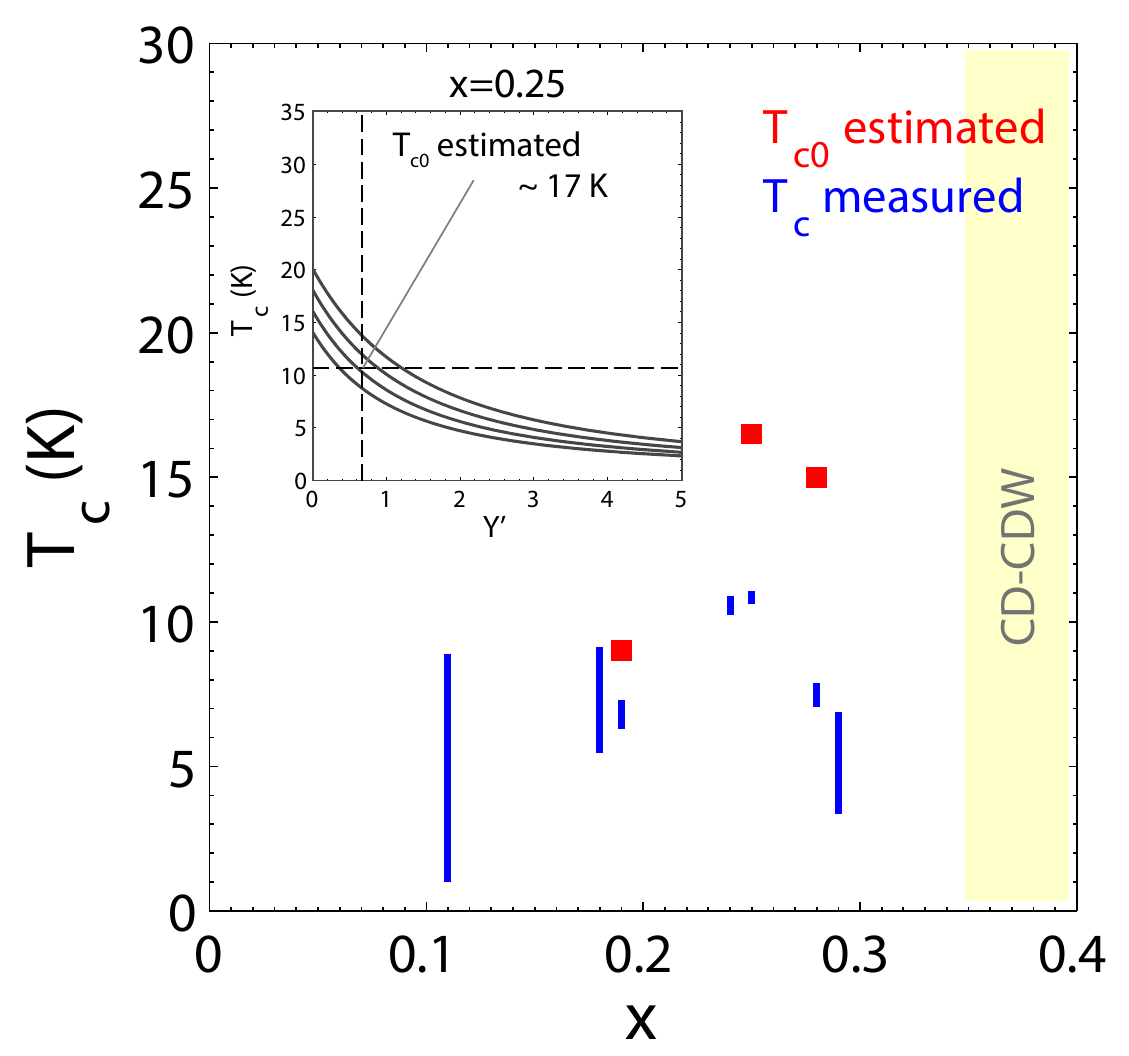}
  \caption{Blue lines correspond to the measured $T_c$ for four to five samples.  Red squares correspond to the inferred $T_{c0}$ when no disorder exits.  Inset shows an example for $x=0.25$ of how $T_c$ decreases with the disorder parameter $Y'$.  From the intersection of the measured $T_c$ and $Y'$ of the material, the disorder-free $T_{c0}$ where $Y'=0$ can be backtracked. }
\label{fig:Tc0_vs_doping_Belitz_Debye145_sq}
\end{figure}

Using this theory, for an assumed value of $T_{c0}$ (or equivalently $\lambda$) and the calculated value of $\mu^*$, we can graphically depict the dependence of $T_c$ on the disorder parameter $Y'$ for BPBO, as shown in the insert of Fig.  \ref{fig:Tc0_vs_doping_Belitz_Debye145_sq} for $x=0.25$.    A family of curves exist for various starting points of $T_{c0}$.  As the disorder parameter $Y'$ increases, $T_c$ is suppressed.  We are able to triangulate which curve in the figure is relevant to our material as we measured $T_c$, whose value is shown by the horizontal dashed line.  We also determined the zero temperature DOS, $N(0)$, from our measurements, hence we know $Y'  = N_F/N(\bar{\omega})-1 = 1/N(0) - 1$, which is shown by the vertical dashed line.   The intersection of these two lines determines the curve relevant to our sample.  If we trace the curve back to  $Y'=0$, then it determines the critical temperature with no disorder, $T_{c0}\approx17$ K for this concentration.  

A similar procedure to obtain the disorder free $T_{c0}$ can be performed for the other two superconducting concentrations for which we obtained the zero temperature DOS.  The estimated $T_{c0}'s$ for  $x = 0.19$, $0.25$, and $0.28$ are $9$, $17$, and $15$ K respectively and are shown as red squares in Fig. \ref{fig:Tc0_vs_doping_Belitz_Debye145_sq}.  The blue lines in this figure shows the measured $T_c$ of four to five samples per composition and was determined from the $0.5$ criteria of $\textrm{d}H_{c2}/\textrm{d}T$ vs. temperature.  The $T_c$'s quoted in this figure are from samples of the same group from which we obtained our samples \cite{GiraldoGallo12}.  One interesting feature of these data is that the disorder-free transition temperatures may not show a peak as a function of $x$.  Uncertainties in the input parameters $(\Theta_D, E_F, n, T_c)$ change $T_{c0}$ at most a few degrees and the uncertainty in $Y'$ is hard to assess in the absence of an understanding of the linear background.  However, the trends observed as $x$ goes to $x_c$ should not be affected.

\begin{figure}[t!]
 \centering
   \includegraphics[width=0.45\textwidth]{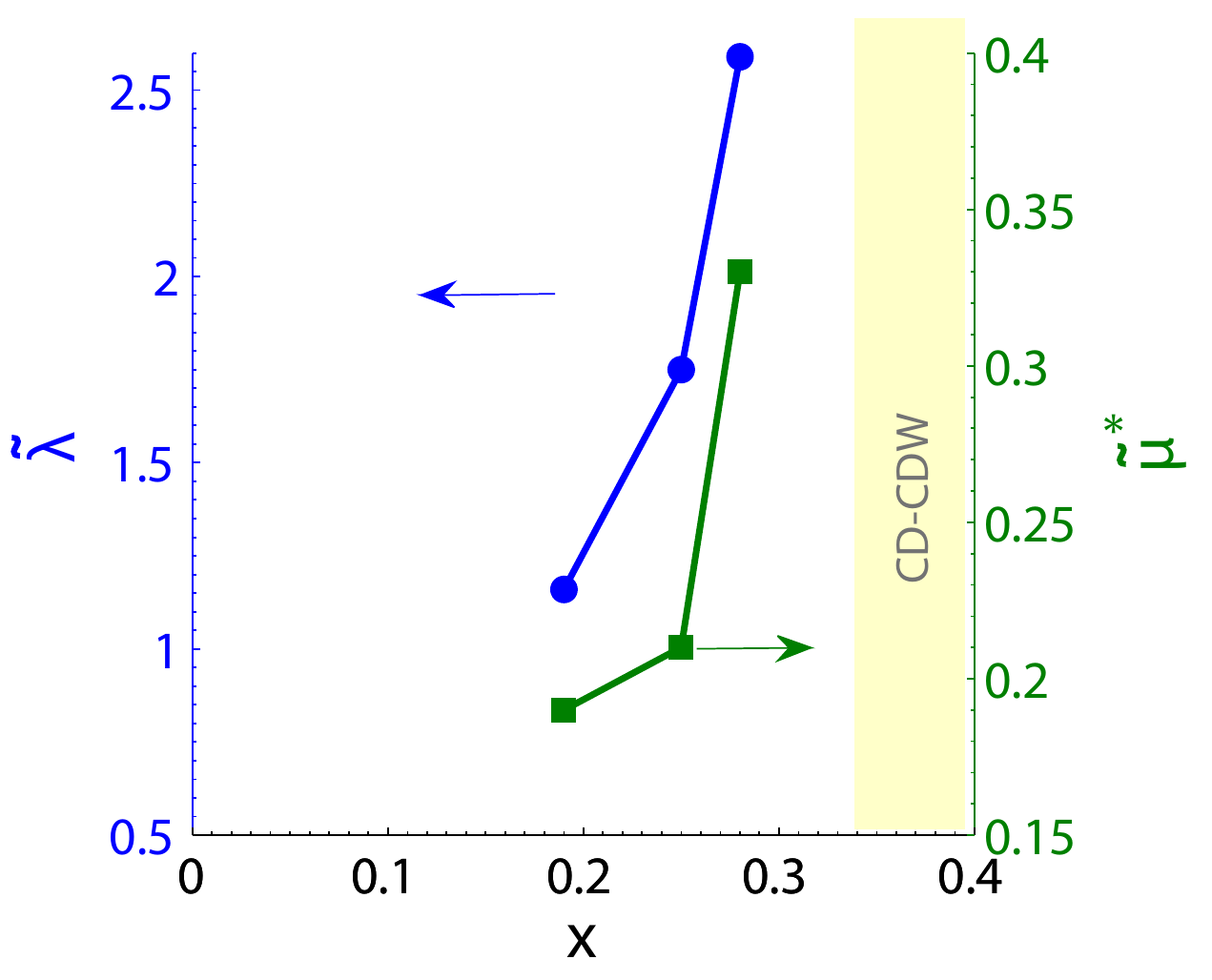}
  \caption{The disorder-dependent electron-phonon coupling strength, $\tilde{\lambda}$, and the disorder-dependent Coulomb pseudopotential, $\tilde{\mu}^*$, is shown as a function of doping, $x$.}
\label{fig:lambdaElPh_deltaMustar_vs_xNoKot_Belitz_tilde}
\end{figure}

In Fig. \ref{fig:lambdaElPh_deltaMustar_vs_xNoKot_Belitz_tilde} we show the corresponding disorder-dependent electron-phonon coupling $\tilde{\lambda}$ (blue circles) and the disorder-dependent Coulomb pseudopotential $\tilde{\mu}$ (green squares) as a function of doping.  We compare the disorder free $\lambda$, which is determined from Eq. \ref{TcBelitz} with $Y'=0$ and $T_c=T_{c0}$,  and the disorder-dependent $\tilde{\lambda}$: $(\lambda, \tilde{\lambda})=(0.94,1.16)$, $(1.45,1.75)$, and $(1.81, 2.03)$ for $x = 0.19$, $0.25$, and $0.28$, respectively.  Similarly, the relationship between the disorder-free $\mu^*$ and the disorder-dependent $\tilde{\mu}^*$ is $(\mu^*, \tilde{\mu}^*)=(0.14,0.19)$, $(0.14, 0.21)$, and $(0.15, 0.28)$ for $x = 0.19$, $0.25$, and $0.28$, respectively.  These results are in accordance with the fact that disorder weakens screening, which effectively increases the Coulomb interactions and electron-phonon coupling. The results also suggest that $\tilde{\lambda}$ and $ \tilde{\mu}^*$ may be diverging as $x$ approaches $x_c$.  This possibility raises interesting theoretical questions.

 It would be interesting to compare these results with BKBO, however, we would need tunneling measurements focusing on the low voltage behavior above $T_c$ , which are not present in the literature to our knowledge.  But we can compare BPBO to BKBO if we use an earlier, less complete version of the effects of disorder on $T_c$ valid for weak coupling and weak disorder \cite{Fukuyama84}.  In this theory, the parameter that characterizes the amount of disorder, $\lambda_{\textrm{loc}}$, can be related to the diffusion constant, which in the dirty limit can be related to upper-critical field measurements $\textrm{d}H_{c2}/\textrm{d}T$ ($\lambda_{\textrm{loc}}=\hbar/2\pi E_F\tau=\hbar/2\pi (1/2 m^* v^2_F)\tau=\hbar/3\pi m^* D\approx(m/3m^*)[(\textrm{d}H_{c2}/\textrm{d}T)/(30\textrm{ kOe/K})]$).  Using upper-critical field measurements, we can again produce similar curves as in Fig. \ref{fig:Tc0_vs_doping_Belitz_Debye145_sq} to obtain $T_{c0}=5$, $11$, $12$, $42$, $52$, and $70$ for  $x=0.11$, $0.19$, $0.20$, $0.24$, $0.25$, and $0.28$ respectively.  At some point we expect the theory of weak disorder and weak coupling to 
  break down with increased concentration, $x$, as a $T_c=70$ K is quite high.  But the trend is the same as seen using the theory of Belitz.

Performing a similar analysis on the optimally doped BKBO, we find that the critical temperature does not change significantly.  For the analysis, we used literature values of the upper critical field $\textrm{d}H_{c2}/\textrm{d}T=-5\pm0.5$ kOe/K and critical temperature $T_c=27$ K \cite{Batlogg88}, the effective mass $m^*=1.5m$ \cite{Puchkov96}, and the Debye temperature $\Theta_D=210$ K \cite{Ott89}.  We took the Fermi energy to be $E_F=1$ eV and the renormalized Coulomb interaction  $\mu^*=0.1$.   We find that while the measured critical temperature is $T_c=27$ K, the disorder-free $T_{c0}=32$ K in the weak-coupling and weak-disorder regime.  This shows that BKBO is relatively unaffected by disorder unlike BPBO.

The origin of disorder in BPBO is currently under investigation by the group of Fisher et. al.  A likely reason for the disorder is due to structural effects in concert with chemical substitution.  The implications of these results in understanding the superconductor-insulator transition with phase fluctuations vs. amplitude effects is currently being investigated.

In summary, we performed PCS measurements on BPBO at various temperatures and for several compositions.  In addition to corroborating results of the superconducting gaps and normal state linear background, we find a disorder driven metal-insulator transition from the characteristic square-root dependence of the conductance at low bias-voltages where the correlation gap disappears around $x_c=0.30$ before the onset of the CD-CDW.  The zero-temperature conductivity as a function of doping also indicates a metal-insulator transition with the same critical concentration of $x_c=0.30$.  We suggest that a scaling theory might be applied to BPBO, though McMillan's scaling theory does not completely fit with our results.  Finally, we estimated the disorder-free critical temperature in BPBO and find that disorder affects the $T_c$ of this material much more greatly than in BKBO.  Our results reconcile the differences seen in the shape of the superconducting dome, as well as $T_c$ values, between BPBO and BKBO and provides a general phase diagram of this family of superconductors.  We hope that our methodology will be useful when assessing the potential of disordered superconductors.

We thank Thomas Devereaux and Phil Wu for useful discussions.  Support for this work came from the Air Force Office of Scientific Research MURI Contract FA9550-09-1-0583-P00006.  KL further acknowledges support from the Lucent Bell Labs Graduate Fellowship.

\bibliographystyle{unsrt}
\bibliography{BPBObib}

\begin{thebibliography}{10}

\bibitem{Belitz89}
D.~Belitz.
\newblock {\em Phys. Rev. B}, 40, 1989.

\bibitem{Fukuyama84}
H.~Fukuyama, H.~Ebisawa, and S.~Maekawa, 1984.

\bibitem{Sleight75}
A.W. Sleight, J.L Gillson, and P.E. Bierstedt.
\newblock {\em Solid State Comm.}, 17, 1975.

\bibitem{Cava88}
R.J. Cava, B.~Batlogg, J.J. Krajewski, R.~Farrow, L.W.~Rupp Jr, A.E. White,
  K.~Short, W.F Peck, and T.~Kometani.
\newblock {\em Nature}, 332, 1988.

\bibitem{Uchida87}
S.~Uchida, K.~Kitazawa, and S.~Tanaka.
\newblock {\em Phase Transitions}, 8, 1987.

\bibitem{Tarapheder96}
A.~Tarapheder.
\newblock {\em Int. J. Mod. Phys. B}, 10, 1996.

\bibitem{Meregalli98}
V.~Meregalli and S.Y. Savrasov.
\newblock {\em Phys. Rev. B}, 57, 1998.

\bibitem{Mattheiss83}
L.F. Mattheiss and D.R. Hamann.
\newblock {\em Phys. Rev. B}, 24, 1985.

\bibitem{Mattheiss85}
L.F. Mattheiss.
\newblock {\em Jap. J. Appl. Phys. Supplement 24-2}, 28, 1983.

\bibitem{Franchini09}
C.~Franchini, G.~Kresse, and R.~Podloucky.
\newblock {\em Phys. Rev. Lett.}, 102, 2009.

\bibitem{Franchini10}
C.~Franchini, A.~Sanna, M.~Marsmann, and G.~Kresse.
\newblock {\em Phys. Rev. B}, 81, 2010.

\bibitem{Yin11}
Z.P. Yin, A.~Kutepov, and G.~Kotliar.
\newblock {\em arXiv:1110.5751}, 2011.

\bibitem{Yin13}
Z.P. Yin, A.~Kutepov, and G.~Kotliar.
\newblock {\em Phys. Rev. X}, 3, 2013.

\bibitem{GiraldoGallo12}
P.~Giraldo-Gallo, H.~Lee, Y.~Zhang, M.J Kramer, M.R. Beasley, T.H Geballe, and
  I.R. Fisher.
\newblock {\em Phys. Rev. B}, 85, 2012.

\bibitem{Ekino89}
Toshikazu Ekino and Jun Akimitsu.
\newblock {\em J. Phys. Soc. Japan}, 58, 1989.

\bibitem{Dynes91PRB}
F.~Sharifi, A.~Pargellis, R.C. Dynes, B.~Miller, E.S. Hellman, J.~Rosamilia,
  and E.H.~Hartford Jr.
\newblock {\em Phys. Rev. B}, 44, 1991.

\bibitem{Szabo97}
P.~Szab\'{o}, P.~Samuely, A.G.M. Jansen, P.~Wyder, J.~Marcus, T.~Klein, and
  C.~Escribe-Filippini.
\newblock {\em J. Of Low Temp Phys.}, 106, 1997.

\bibitem{Suzuki92}
Morio Suzuki, Kazunori Komorita, Hiroshi Nakano, and Leo Rinderer.
\newblock {\em J. Phys. Soc. Japan}, 61, 1992.

\bibitem{Sharifi91}
F.~Sharifi, A.~Pargellis, and R.C. Dynes.
\newblock {\em Phys. Rev. Lett.}, 67, 1991.

\bibitem{Hertel83}
G.~Hertel, D.J. Bishop, E.G. Spencer, J.M Rowell, and R.C. Dynes.
\newblock {\em Phys. Rev. Lett.}, 50, 1983.

\bibitem{Altshuler79}
B.L Altshuler and A.G. Aronov.
\newblock {\em Solid State Comm.}, 30, 1979.

\bibitem{McMillan81}
W.L. McMillan.
\newblock {\em Phys. Rev. B}, 24, 1981.

\bibitem{Lee82}
P.A. Lee.
\newblock {\em Phys. Rev. B}, 26, 1982.

\bibitem{Lee85}
Patrick~A. Lee and T.V. Ramakrishnan.
\newblock {\em Rev. Mod. Phys.}, 57, 1985.

\bibitem{Itoh84}
T.~Itoh, K.~Kitazawa, and S.~Tanaka.
\newblock {\em J. Phys. Soc. Japan}, 53, 1984.

\bibitem{Tani80}
T.~Tani, T.~Itoh, and S.~Tanaka.
\newblock {\em J. Phys. Soc. Jpn. Suppl. A}, 49.

\bibitem{Thanh80}
T.D. Thanh, A.~Koma, and S.~Tanaka.
\newblock {\em Appl. Phys.}, 22, 1980.

\bibitem{Kitazawa85}
K.~Kitazawa, M.~Naito, T.~Itoh, and S.~Tanaka.
\newblock {\em J. Phys. Soc. Jpn.}, 54, 1985.

\bibitem{Morel62}
P.~Morel and P.W. Anderson.
\newblock {\em Phys. Rev. B}, 12, 1962.

\bibitem{Batlogg88}
B.~Batlogg, R.J. Cava, L.W.~Rupp Jr., A.M. Mujsce, J.J. Krajewski, J.P.
  Remeika, and W.F.~Peck Jr.
\newblock {\em Phys. Rev. Lett.}, 61, 1988.

\bibitem{Puchkov96}
A.V. Puchkov, T.~Timusk, M.A. Karlow, S.L. Cooper, P.D. Han, and D.A. Payne.
\newblock {\em Phys. Rev. B}, 54, 1996.

\bibitem{Ott89}
K.C. Ott, M.F. Hundley, G.H. Kwei, M.P. Maley, M.E. McHenry, E.J. Peterson,
  J.D. Thompson, and J.O. Willis.
\newblock {\em Mat. Res. Sco. Symp. Proc.}, 156, 1989.

\end{thebibliography}
\end{document}